\documentclass[nofootinbib,superscriptaddress,10pt,aps,prd,preprint]{revtex4-1}
\usepackage[utf8]{inputenc}
\usepackage{amsmath,mathtools}

\usepackage{tensor}
\usepackage{amssymb}
\usepackage{graphicx}
\usepackage{physics}
\usepackage{lipsum}  
\usepackage{bigints}
\usepackage{verbatim}

\usepackage{braket}
\usepackage{cancel}
\usepackage{booktabs}
\usepackage{color}
\usepackage{hyperref}
\newcommand{\be}{\begin{equation}}
\newcommand{\ee}{\end{equation}}
\newcommand{\bea}{\begin{eqnarray}}
\newcommand{\eea}{\end{eqnarray}}

\newcommand{\non}{\nonumber}

\newcommand{\A}{\mathcal{A}_{\pm}}
\newcommand{\Aone}{\pm i \xi_1}
\newcommand{\Bone}{\nu_1+\frac{1}{2}}
\newcommand{\Atwo}{\mp i \xi_2}
\newcommand{\Btwo}{\nu_2-\frac{1}{2}}
\newcommand{\zi}{2ik\eta_i}
\newcommand{\z}{2ik\eta}
\DeclareMathOperator{\arcsinh}{arcsinh}

\DeclareMathOperator{\arctanh}{arctanh}

\begin{document}
\title{Inflationary helical magnetic fields with a sawtooth coupling}
\author{Chiara Cecchini,}
\email{chiara.cecchini@unitn.it}
\author{Massimiliano Rinaldi.}
\email{massimiliano.rinaldi@unitn.it}
\affiliation{Dipartimento di Fisica, Università di Trento,\\Via Sommarive 14, I-38123 Povo (TN), Italy}
\affiliation{Trento Institute for Fundamental Physics and Applications TIFPA-INFN,\\Via Sommarive 14, I-38123 Povo (TN), Italy}

\begin{abstract}
We study the generation of helical magnetic fields during inflation by considering a model which does not suffer from strong coupling or large back-reaction. Electromagnetic conformal invariance is broken only during inflation by coupling the gauge-invariants $F_{\mu\nu}F^{\mu\nu}$ and $F_{\mu\nu}\tilde{F}^{\mu\nu}$ to a time-dependent function $I$ with a sharp transition during inflation. 
The magnetic power spectrum is scale-invariant up to the transition and very blue-tilted after that. 
The subsequent evolution of the helical magnetic field is subjected to magneto-hydrodynamical processes, resulting in far larger coherence lengths than those occurring after adiabatic decay.
Scale-invariant quadratic gravity is a suitable framework to test the model, providing a natural physical interpretation. We show that fully helical magnetic fields are generated with values in agreement with the lower bounds on fields in the Intergalactic Medium derived from blazar observations. This model holds even at large/intermediate energy scales of inflation, contrary to what has been found in previous works.
\end{abstract}

\maketitle

\section{Introduction}
\label{sect:1}
Magnetic fields are ubiquitous in the Universe. 
Galaxies and clusters of galaxies host magnetic fields with a strength from few to tens of $\mu$G and coherent on scales of tens kpc, independently on their redshift \cite{bernet_2008, feretti_2012, beck_2013, digennaro_2021}. 
Further, gamma-ray observations of distant blazars have inferred the presence of magnetic fields with large coherence lengths $L\gtrsim$1 Mpc in the cosmic voids of the intergalactic medium (IGM)  \cite{neronov_2010, taylor_2011}. Together with upper bounds from CMB observations \cite{planckcollaboration_2016, sutton_2017, giovannini_2018, paoletti_2022} and ultra-high-energy cosmic rays \cite{bray_2018, neronov_2021}, they constrain the amplitude of IGM magnetic fields coherent on Mpc scales or larger as $10^{-16}\lesssim B_0\lesssim 10^{-9}$ G. The lower bound is larger by a factor $(L/1\, \mathrm{Mpc})^{-1/2}$ for $L<$1 Mpc \cite{neronov_2010}. 
Moreover, non-vanishing parity-odd correlators of gamma ray arrival directions observed by Fermi-LAT suggest that intergalactic magnetic fields are helical \cite{tashiro_2013, tashiro_2014, chen_2015}. 
For updated reviews on observational constraints, see, for instance, \cite{paoletti_2019, vachaspati_2021}.

If, on the one hand, small-scale magnetic fields could be explained in terms of purely astrophysical processes,  e.g., Biermann battery and subsequent amplification via adiabatic contraction and cosmic dynamos \cite{durrer_2013, brandenburg_2005}, on the other, a primordial magnetogenesis is favorite to explain IGM fields, too. 
Several scenarios have been considered (for reviews, see \cite{subramanian_2016, grasso_2001, widrow_2012, kandus_2011}). 
Cosmological phase transitions -- provided they can be treated as first-order ones -- lead to quite discouraging limits on causally-generated magnetic fields since their coherence length is bounded by the horizon size during the transition \cite{vachaspati_1991}. The amplitudes are too small to seed dynamo even when helicity is included \cite{caprini_2009}.  
For this reason, inflation seems the ideal setting for early-time magnetogenesis by naturally providing macroscopic coherence scales. As for primordial scalar and tensor perturbations, quantum vacuum fluctuations of the gauge field could have been amplified and converted into observable electromagnetic (EM) fields. 
Nevertheless, the conformal invariance of standard electromagnetism prevents the magnetic field from amplification in a spatially flat Friedmann-Lemaitre-Robertson-Walker (FLRW) background, washing out its amplitude as $B\sim a^{-2}$.

As a remark, the above conclusion holds only in a spatially flat FLRW Universe. When instead an open geometry is considered, the FLRW spacetime is conformally flat only locally, and it was argued \cite{barrow_2011, barrow_2012} that it could lead to a slowing down of the adiabatic decay rate of magnetic fields. However, as extensively discussed in \cite{shtanov_2013, adamek_2012, yamauchi_2014}, the resulting magnetic field amplitude is unlikely to be of astrophysical significance. Therefore, inflationary magnetogenesis requires a breaking of conformal symmetry in the EM sector \cite{turner_1988}.

Since mechanisms that violate gauge invariance are known to bring ghost instabilities \cite{himmetoglu_2009}, most of the attention has been devoted to gauge-invariant models of magnetogenesis.
The most commonly studied cases involve non-minimal couplings of the EM field to gravity $\propto R^n F_{\mu\nu}F^{\mu\nu}$ \cite{mazzitelli_1995}, or time-dependent couplings $\propto I^2(t)F_{\mu\nu}F^{\mu\nu}$. The latter, first studied by Ratra \cite{ratra_1992}, can explain large enough magnetic field strength and coherence lengths. Still, they typically suffer either from large back-reaction of the EM field -- when $I$ decreases monotonically with scale factor $a$ -- or strong coupling \cite{demozzi_2009}, in the opposite case. Further modifications of the Ratra model have been proposed to solve these problems \cite{ferreira_2013, campanelli_2015, sharma_2017, sharma_2018, nandi_2021}. For example, combining an increasing and a decreasing function avoids both problems. This comes to the price of significantly lowering the energy scale of inflation or the reheating temperature, depending on the model, resulting in lower magnetic field amplitudes. 
Another possibility is the axion model \cite{garretson_1992}, which considers a Lagrangian density $\mathcal{L}\propto - F_{\mu\nu}F^{\mu\nu} - g_{\phi\gamma}\phi F_{\mu\nu}\tilde{F}^{\mu\nu}$, with $\phi$ a rolling pseudoscalar and $g_{\phi\gamma}$ its coupling constant to photons. Even though the axion model provides too small magnetic field amplitudes, it was recently reconsidered since the introduction of the parity-violating term in the action leads to maximally helical magnetic fields \cite{anber_2006}. When helical fields are considered, non-linear evolution processes like inverse cascade become relevant, leading to larger coherence lengths than predicted by adiabatic decay in the non-helical case \cite{son_1999, field_2000, christensson_2001}. 
These results have motivated the study of a mixed axion-Ratra model of magnetogenesis, first presented in \cite{caprini_2014}, further investigated in  \cite{caprini_2018}, and generalized in \cite{cheng_2014}. This hybrid model contains both couplings, i.e. $\propto I^2(t)F_{\mu\nu}F^{\mu\nu}$ and $\propto I^2(t)F_{\mu\nu}\tilde{F}^{\mu\nu}$, where $I$ is a monotonic decreasing function during inflation. It can generate a blue-tilted magnetic field with $B_0 \lesssim 10^{-21}\, \mathrm{G}$ at Mpc scales, provided the energy scale of inflation is significantly lowered from its upper bound, which is $\rho_{inf}^{1/4}< 1.6 \times 10^{16}\, \mathrm{GeV}$ according to \textit{Planck} \cite{planckcollaboration_2020Inflation}. 

In this paper, we modify the hybrid axion-Ratra model by taking a non-monotonic function $I(t)$ with a sharp transition. The function $I$ is modeled as a power law of the scale factor, this being the most natural and analytically tractable choice. In this model, $I\neq1 $ only during inflation, so this is the only period where EM conformal invariance is broken. The strong coupling problem is avoided by choosing $I>1$ during inflation. At the same time, thanks to the sawtooth shape of the function $I$, the back-reaction problem affects only the second stage of inflation, corresponding to the decreasing branch of $I$. Consequently, viable magnetogenesis could arise even at large and intermediate inflationary energy scales, resulting in larger magnetic field amplitudes. Moreover, macroscopic correlation lengths are predicted because the inverse cascade mechanism significantly affects the post-inflationary evolution of fully helical fields. 
Other works studied the generation of helical magnetic fields with a non-monotonic coupling $I$ \cite{sharma_2018, fujita_2019}, but to our knowledge this has only been done in the context of a mixed inflation-reheating magnetogenesis. For this reason, the analytical results derived here can be useful, as they are valid in any inflationary background. 

It is sometimes claimed \cite{durrer_2022} that such phenomenologically designed coupling functions are hardly justified without physical motivations, especially if they involve sharp transitions. For this reason, we apply our model to scale-invariant quadratic gravity as first presented in \cite{rinaldi_2016} and further examined in \cite{tambalo_2017, vicentini_2019, ghoshal_2022}. This model is compatible with an arbitrarily long inflationary phase and it predicts spectral indices in agreement with \textit{Planck} data \cite{planckcollaboration_2020Inflation}. We will show that it is possible to find a function of the inflaton field reproducing the sawtooth coupling $I$ to a good approximation. Associating $I$ to the inflaton's evolution provides an intuitive physical interpretation of this model of magnetogenesis.

The analysis of magnetogenesis presented in this paper does not take into account the dynamics of the stochastic noises of EM perturbations, as recently investigated in \cite{talebian_2020, talebian_2022}. 
It was shown \cite{talebian_2020} that stochastic effects can enhance the magnetic field amplitude significantly for the standard Ratra coupling, leading to $B_0 \sim 10^{-13}\, \mathrm{G}$ on Mpc scales. On the other hand, when the axion coupling is included, the resulting amplitudes are not sufficient to provide a chiral primordial seed for galactic magnetic fields, nor for IGM ones \cite{talebian_2022}. The stochastic approach, being complementary to the one presented here, could be applied also to this model of magnetogenesis. 

The paper is organized as follows. Sec. \ref{sect:2} is entirely devoted to the model of magnetogenesis: we compute solutions to the modified Maxwell equations and the resulting EM power spectra. Here, also the evolution of the magnetic field after inflation is discussed, following the literature \cite{caprini_2014}. In Sec. \ref{sect:3} we briefly present the main features of the scale-invariant model of quadratic gravity proposed in \cite{rinaldi_2016} and we find an approximate expression for the inflaton's evolution. In Sec. \ref{sect:4} we provide some significant results for current values of magnetic field amplitude and coherence length, for different energy scales of inflation. We further verify that both back-reaction and overproduction of gravitational waves by the gauge field are avoided with the adopted choice of parameters.
The possibility of generating a baryon asymmetry from decaying magnetic helicity, as studied in \cite{kamada_2016}, is also considered. The final Sec. \ref{sect:5} is devoted to a discussion on our results and a summary of the work. 

In this paper, we adopt a spatially flat FLRW metric, $ds^2=-dt^2+a^2(t)\delta_{ij}dx^i dx^j$ that we mostly use written in conformal time $\eta$, namely $ds^2=-a^2(\eta)[\eta_{\mu\nu}dx^\mu dx^\nu]$.  We further set $\hbar = c = 1$. $M_{pl} = m_{pl}/\sqrt{8\pi}$ is the reduced Planck mass. Subscripts $i, f$ denote the beginning and end of inflation, respectively, while $*$ identifies the transition point. 
\section{Helical magnetic fields from a sawtooth coupling}
\label{sect:2}
To generate helical magnetic fields during inflation, we consider a hybrid of the Ratra and the axion model, as first proposed in \cite{caprini_2014}. We further assume that the time dependence that explicitly breaks conformal invariance comes from a coupling $I$ between the gauge fields and a generic scalar field, $\phi$. Therefore we consider
\be
\label{eq:generalaction}
    S = -\dfrac{1}{16\pi}\int d^4x\sqrt{-g}\,  I^2(\phi)\left(F_{\mu\nu}F^{\mu\nu} - \gamma F_{\mu\nu}\tilde{F}^{\mu\nu}\right) + \int d^4x\sqrt{-g}\, \mathcal{L}_{\phi}, 
 \ee
where $F_{\mu\nu} = \partial_{\mu}A_{\nu} - \partial_{\nu}A_{\mu}$ is the field strength of a $U(1)$ gauge field and $\tilde{F}^{\mu\nu} = (1/2)\epsilon^{\mu\nu\alpha\beta}F_{\alpha\beta}$ is its dual with the totally antisymmetric tensor $\epsilon^{\mu\nu\alpha\beta}$ defined as $\epsilon^{\mu\nu\alpha\beta} = 1/\sqrt{-g}\,\eta^{\mu\nu\alpha\beta}$. $\eta^{\mu\nu\alpha\beta}$ is the Levi-Civita symbol with values $\pm 1$. $\gamma$ is a positive dimensionless constant. Note that we did not include a source term $\sim j^{\mu}A_{\mu}$ since we assume negligible free charge density during inflation. 
The gauge field is assumed to be a test field with no influence on the scalar field evolution or spacetime geometry. This statement will be verified with a detailed analysis on back-reaction. For the time being, we do not make any assumption on the evolution of the scalar field given by $\mathcal{L}_{\phi}$, i.e., on the inflationary dynamics. The calculations to derive the EM spectra, therefore, follow the literature (for extensive reviews, see \cite{subramanian_2010, subramanian_2016}). In the following, we summarize the main results. 

The gauge freedom is fixed by adopting the Coulomb gauge, $A_0 = \partial_iA^i = 0$. We decompose and quantize the gauge field as 
\be
    A^i(\mathbf{x}, \eta) = \sqrt{4\pi}\int \dfrac{d^3k}{(2\pi)^3}\sum_{\lambda = \pm} \varepsilon^i_{\lambda}(\mathbf{k}) \left[b_{\lambda}(\mathbf{k})A_{\lambda}(k, \eta)e^{i\mathbf{k}\cdot\mathbf{x}} + b_{\lambda}^{\dagger}(\mathbf{k})A^*_{\lambda}(k, \eta)e^{-i\mathbf{k}\cdot\mathbf{x}}\right], 
\ee
where $\eta$ is the conformal time and \textbf{k} is the comoving wave vector, related as $\mathbf{k} = a\mathbf{k}_{ph}$ to the physical wave vector $\mathbf{k}_{ph}$. $\varepsilon^i_{\lambda}$ is the helicity vector associated to the helicity state $\lambda = \pm$ \cite{caprini_2014, caprini_2018}. 
$b_{\pm}(\mathbf{k}), b_{\pm}^{\dagger}(\mathbf{k})$ are the creation and annihilation operators obeying the usual commutation relations, $\left[b_{\lambda}(\mathbf{k}), b^{\dagger}_{\sigma}(\mathbf{k'})\right] = (2\pi)^3\delta(\mathbf{k} - \mathbf{k'})\delta_{\lambda\sigma}$.

Variations of \eqref{eq:generalaction} with respect to $A_{\mu}$ lead to the Maxwell equations on the spatially flat FLRW metric with the coupling $I$. In the Fourier space, these are
\be
\label{eq:Maxwellgeneral}
    \A''(k, \eta) + \left(k^2 \pm 2\gamma k \dfrac{I'}{I}-\dfrac{I''}{I}\right)\A(k, \eta)=0, 
\ee
where we have introduced the canonically normalized field $\A = I(\eta) A_{\pm}(k, \eta)$. The prime denotes a derivative with respect to conformal time.
The non-helical case is recovered by setting $\gamma = 0$. Solutions to \eqref{eq:Maxwellgeneral} can be found under specific assumptions on the coupling function $I$. A monotonic coupling $I\sim a^{n}$ was considered in \cite{caprini_2014}; there, it was shown that, by choosing $\gamma\sim\mathcal{O}(10)$, magnetic fields compatible with lower bounds on IGM fields could be generated without strong coupling or back-reaction problems. This scenario is possible provided inflation occurs at an energy scale between $10^5$ and $10^{10}$ GeV.

In this work, we consider a sawtooth coupling function, first investigated for the non-helical case in \cite{ferreira_2013} and later studied in a mixed inflation-reheating magnetogenesis in \cite{sharma_2017, sharma_2018}. One motivation is to provide a fully inflationary magnetogenesis mechanism for any sensible inflationary energy scale. 

\subsection{Sawtooth coupling}
It is known that $I$ as written in \eqref{eq:generalaction} plays the role of an inverse coupling constant \cite{demozzi_2009}. Therefore, the coupling function should satisfy $I \gtrsim 1$ throughout inflation and approach unity in the end, thereby avoiding strong coupling and restoring conformal invariance once inflation is over. We consider the possibility of having a sharp transition in the coupling function, which can be parameterized as
\be
\label{eq:coupling}
    I = 
    \begin{cases} 
       \mathcal{C}\left(\dfrac{a}{a_*}\right)^{\nu_1}& a_i<a<a_*\\
       \mathcal{C}\left(\dfrac{a}{a_*}\right)^{-\nu_2}& a_*<a<a_f 
    \end{cases}
\ee
where $\nu_i$ are positive exponents. In the following, we will refer to $a_i < a < a_*$ as \textit{first stage} and $a_* < a < a_f$ as \textit{second stage} of inflation.
The constant $\mathcal{C}$ and the scale factor at transition $a_*$ are found by imposing $I_i = I_f = 1$; in this way, EM conformal invariance is only broken during inflation. Note that this condition cannot be achieved for monotonic coupling functions.

We can solve \eqref{eq:Maxwellgeneral} for three regimes: in the short wavelength limit (vacuum solutions), during the first stage, and during the second stage. We examine the three cases separately. 
\subsubsection{Vacuum solutions}
The initial conditions are specified for each mode (or wavenumber $k$) in the short wavelength limit, $(-k\eta)\gg 1$, where equation \eqref{eq:Maxwellgeneral} reduces to the usual wave equation: 
\be
\label{Eq:Maxwellvacuum}
    \mathcal{A}''(k, \eta) + k^2 \mathcal{A}(k, \eta) = 0. 
\ee
Being deep within the Hubble sphere, the modes $\mathcal{A}$ are not amplified, thus we can solve \eqref{Eq:Maxwellvacuum} with the Bunch-Davies vacuum conditions, 
\be
    \mathcal{A}^{BD}(k, \eta) = \dfrac{1}{\sqrt{2k}}e^{-ik(\eta -\eta_i)}, 
\ee
where the constant phase $e^{ik\eta_i}$ is added for later convenience. 
\subsubsection{First stage}
When conformal invariance is explicitly broken by the coupling $I$ and before transition, equation $\eqref{eq:Maxwellgeneral}$ becomes
\be
\label{eq:Maxwellfirst}
    \A''(k, \eta) + \left(k^2 \mp 2 k \dfrac{\xi_1}{\eta} -\dfrac{\nu_1(\nu_1+1)}{\eta^2}\right)\A(k, \eta)=0, \qquad \xi_1 = \gamma\nu_1, 
\ee
where we have considered a purely de Sitter expansion, $\eta = -1/aH$. As we will see, the parameter $\xi_1 = \gamma \nu_1$ quantifies the enhancement of the magnetic field amplitude \cite{caprini_2014}. The general solutions can be written as a linear combination of the Whittaker functions $M_{\alpha, \beta}(z)$ and $W_{\alpha, \beta}(z)$ \cite{abramowitz_1988}, 
\be
    \A^I(k, \eta) = \dfrac{1}{\sqrt{2k}}\left[C_1^{\pm} M_{\Aone,\Bone}(\z) + C_2^{\pm} W_{\Aone, \Bone}(\z)\right].
\ee
The integration constants $C_i^{\pm}$ are found by imposing the following junction conditions \cite{ferreira_2013}: 
\be
    \dfrac{\mathcal{A}^{BD}(k, \eta_i)}{I_i} = \dfrac{\A^{I}(k, \eta_i)}{I_i}, \qquad \left.\dfrac{\partial}{\partial \eta}\left(\dfrac{\mathcal{A}^{BD}}{I}\right)\right|_{\eta = \eta_i} = \left.\dfrac{\partial}{\partial \eta}\left(\dfrac{\A^I}{I}\right)\right|_{\eta = \eta_i}.
\ee
Note that the factor $I_i = 1$ at the denominator in the first equation has been written to stress that we require the continuity of the physical vector potential $A_{\pm}(k, \eta)$, not of the canonically normalized one. The system can be solved exactly to give\footnote{We employed the identity: $M_{\alpha, \beta}(z)\partial_z W_{\alpha, \beta}(z) - W_{\alpha, \beta}(z)\partial_z M_{\alpha, \beta}(z) = - \Gamma(2\beta + 1)/\Gamma(\beta-\alpha+1/2)$}
\bea
    &C_1^{\pm}& = \dfrac{\Gamma(\nu_1 + 1\mp i \xi_1)}{\zi\Gamma(2\nu_1 + 2)}\left[W_{1\Aone, \Bone}(\zi) - \left(\zi + \nu_1 \mp i\xi_1\right)W_{\Aone, \Bone}(\zi)\right],\\
    &C_2^{\pm}& = \dfrac{\Gamma(\nu_1 + 1\mp i \xi_1)}{\zi\Gamma(2\nu_1 + 2)}\Bigl[\left(\zi + \nu_1 \mp i \xi_1\right)M_{\Aone, \Bone}(\zi)+ \\&&\hspace{3.7cm}+ \left(1+\nu_1\Aone\right)M_{1\Aone, \Bone}(\zi)\Bigr].\non
\eea
For later convenience, we stress that, in the sub-horizon limit $(-k\eta_i)\gg 1$, $|C_1^{\pm}|\to 0$ as $(-k\eta_i)^{-1}$, while $|C_2^{\pm}|\simeq e^{\pm \frac{\pi}{2}\xi_1 }$, reflecting the effect of the parity violating term introduced in the action \eqref{eq:generalaction}. In other words, only the `positive' polarization mode $(\propto C_2^{+})$ is amplified, while the `negative' one is exponentially suppressed.
\subsubsection{Second stage}
After transition, equation $\eqref{eq:Maxwellgeneral}$ becomes
\be
\label{eq:Maxwellsecond}
    \A''(k, \eta) + \left(k^2 \pm 2 k \dfrac{\xi_2}{\eta} -\dfrac{\nu_2(\nu_2-1)}{\eta^2}\right)\A(k, \eta)=0, \qquad \xi_2 = \gamma\nu_2, 
\ee
with general solution of the form
\be
    \A^{II}(k, \eta) = \dfrac{1}{\sqrt{2k}}\left[D_1^{\pm} M_{\Atwo,\Btwo}(\z) + D_2^{\pm} W_{\Atwo, \Btwo}(\z)\right].
\ee
The junction conditions are now imposed at the moment of transition $\eta = \eta_*$:
\be
\label{eq:Junction2}
    \dfrac{\A^{I}(k, \eta_*)}{I_*} = \dfrac{\A^{II}(k, \eta_*)}{I_*}, \qquad \left.\dfrac{\partial}{\partial \eta}\left(\dfrac{\A^{I}}{I}\right)\right|_{\eta = \eta_*} = \left.\dfrac{\partial}{\partial \eta}\left(\dfrac{\A^{II}}{I}\right)\right|_{\eta = \eta_*}.
\ee
The expressions for $D^{\pm}_i$ are rather lengthy, so by following \cite{fujita_2019} we just report their expressions in the super-Hubble limit $(-k\eta_*)\ll1$
\bea
    &D_1^{\pm}& \simeq C_2^{\pm}\dfrac{\Gamma(1+2\nu_1)}{\Gamma(1+\nu_1\mp i\xi_1)}\dfrac{1}{(2ik\eta_*)^{\nu_1+\nu_2}},\\
    &D_2^{\pm}& \simeq C_2^{\pm}\dfrac{\Gamma(2\nu_1-1)\Gamma(\nu_2\pm i\xi_2)}{\Gamma(2\nu_2)\Gamma(\nu_1\mp i \xi_1)}\dfrac{(1\pm i \gamma)(\nu_1+\nu_2)}{1+2\nu_2}\left(2ik\eta_*\right)^{1-\nu_1+\nu_2}.
\eea

\subsection{Power spectra}
Once the analytical expressions for $\A^{I, II}$ are known, we can compute the electromagnetic power spectra and their evolution in time. 
We introduce the electric and magnetic energy density per logarithmic interval in $k$ space for each polarization mode as \cite{subramanian_2016}
\be
    \dfrac{{d\rho_B^{\pm}}}{d\ln k} = \dfrac{1}{2\pi^2}\left(\dfrac{k}{a}\right)^4k \left|\A\right|^2, \qquad \dfrac{d\rho_E^{\pm}}{d\ln k} = \dfrac{I^2}{2\pi^2}\dfrac{k^3}{a^4}\left|\left[\dfrac{\A(k, \eta)}{I}\right]'\right|^2.
\ee
By taking the asymptotic expansion of the Whittaker functions, we can compute the spectra in the super-Hubble limit, in both stages,
\bea
    \label{eq:PE1}&\dfrac{d\rho_{E, I}^{\pm}}{d\ln k}& \simeq \mathcal{F}_{E, I}^{\,\pm}(\nu_1) H^4 (-k\eta)^{4-2\nu_1},\\
    &&\quad\mathcal{F}_{E, I}^{\,\pm}(\nu_1) = e^{\pm \pi \xi_1}\dfrac{\xi_1^2}{2^{2\nu_1}\pi^2}\left|\dfrac{\Gamma(2\nu_1)}{\Gamma(1+\nu_1\mp i \xi_1)}\right|^2,\non\\
    \label{eq:PB1}&\dfrac{d\rho_{B, I}^{\pm}}{d\ln k}& \simeq \mathcal{F}_{B, I}^{\,\pm}(\nu_1)H^4 (-k\eta)^{4-2\nu_1},\\
    &&\quad \mathcal{F}_{B, I}^{\,\pm}(\nu_1) = \mathcal{F}_{E, I}^{\,\pm}(\nu_1),\non\\
    \label{eq:PE2}&\dfrac{d\rho_{E, II}^{\pm}}{d\ln k}& \simeq \mathcal{F}_{E, II}^{\,\pm}(\nu_1, \nu_2) H^4(-k\eta)^{6-2\nu_1}\left(\dfrac{\eta_*}{\eta}\right)^{2-2\nu_1+2\nu_2},\\
    &&\quad \mathcal{F}_{E, II}^{\,\pm}(\nu_1, \nu_2) = e^{\pm \pi \xi_1}\dfrac{2^{2-2\nu_1}}{\pi^2}\left|\dfrac{\Gamma(2\nu_1-1)(1\pm i \gamma)(\nu_1+\nu_2)}{\Gamma(\nu_1\mp i \xi_1)(1+2\nu_2)}\right|^2\non,\\
    \label{eq:PB2}&\dfrac{d\rho_{B, II}^{\pm}}{d\ln k}& \simeq \mathcal{F}_{B, II}^{\,\pm}(\nu_1, \nu_2)H^4(-k\eta)^{8-2\nu_1}\left(\dfrac{\eta_*}{\eta}\right)^{2-2\nu_1+2\nu_2},\\
    &&\quad \mathcal{F}_{B, II}^{\,\pm}(\nu_1, \nu_2) = e^{\pm \pi\xi_1}\dfrac{2^{2-2\nu_1}}{\pi^2}\left|\dfrac{\Gamma(2\nu_1-1)\Gamma(2\nu_2-1)(1\pm i\gamma)(\nu_1+\nu_2)}{\Gamma(2\nu_2)\Gamma(\nu_1\mp i\xi_1)(1+2\nu_2)}\right|^2\non.
\eea
The total electric and magnetic spectrum in each stage is the sum of the contributions coming from both polarization modes. 
Note that we have employed the sub-horizon limit for the coefficients $C_2^{\pm}$. A few comments are in order here: in the first stage of inflation, the scale invariance condition for the magnetic power spectrum is attained for $\nu_1 = 2$. There, the property of scale invariance and having an $\eta$-independent $\rho^{\pm}_{B, I}$ go together \cite{subramanian_2016}. Most interestingly, when $\nu_1 = 2$ also the electric power spectrum, having the same shape of the magnetic one, is scale-invariant: this result is a consequence of having a monotonic increasing coupling in the first stage -- which in turn is only possible for some sawtooth coupling\footnote{The conditions to avoid strong coupling and to recover conformal invariance at the end are $I(a) \geq1$ and $I_f = 1$, respectively. Therefore, the only allowed monotonic coupling $I$ is a decreasing function of the scale factor.} -- and having helicity since such a behavior does not show for a sawtooth coupling without helicity \cite{sharma_2017, ferreira_2013}. In the second stage of inflation, on the other hand, when $\nu_1 = 2$ the electromagnetic spectra are blue-tilted, namely $d\rho^{\pm}_{E, II}/d\ln k \sim k^2$ and $d\rho^{\pm}_{B, II}/d\ln k \sim k^4$. This means that, due to the matching procedure \eqref{eq:Junction2}, it is impossible to have a scale-independent magnetic power spectrum throughout inflation \cite{ferreira_2013}.
Finally, note that the $k$- and $\eta$-dependence are decoupled in the spectra of the second stage: having a scale-invariant spectrum does not necessarily imply that the spectrum is also $\eta$-independent. 

\subsection{Cosmological evolution of magnetic fields}
In the following, we introduce the relevant quantities to compute the observables of the current magnetic field and to verify the absence of back-reaction. The energy stored in the EM field at a given time $\eta = -1/aH$ is given by
\be
\label{eq:rhoEM}
    \rho_{EM} (a) = \rho_E(a) + \rho_B(a) = \int_{k_i = a_i H}^{k = aH} \dfrac{dk}{k}\left(\dfrac{d\rho^{\pm}_E}{d\ln k} + \dfrac{d \rho^{\pm}_B}{d \ln k}\right), 
\ee
where summation over the two polarization modes is implicit. In writing the extremes of integration, we have considered that at the moment $\eta_k$, when the corresponding wave crosses the Hubble scale, the scale factor is $a_k \simeq k/H$ \cite{demozzi_2009}. According to the time $\eta$ of interest, we will consider the solution for the power spectrum in the first or the second stage of inflation from equations \eqref{eq:PE1}-\eqref{eq:PB2}, as explained in Appendix \ref{appendixA}.

In agreement with \cite{durrer_2013}, we define the comoving characteristic scale of the magnetic field, which is sometimes referred to as ``correlation scale'', 
\be
\label{eq:correlationscale}
    L_c = \dfrac{2\pi}{\rho^{\pm}_B}\int \dfrac{dk}{k^2} \dfrac{d\rho^{\pm}_B}{d\ln k}. 
\ee
Here again, the integral is performed over the $k$-range of interest at a given time. We further define the ``characteristic'' magnetic field strength at scale $\lambda = 2\pi/k$ as
\be
\label{eq:B}
    B_{\lambda} = \left.\sqrt{\dfrac{8\pi}{I^2}\dfrac{d\rho^{\pm}_B}{d\ln k}}\right|_{k = 2\pi/\lambda}, 
\ee
and the corresponding scale-averaged value
\be
\label{eq:cascade1}
    B = \sqrt{\dfrac{8\pi}{I^2}\rho^{\pm}_B}\,. 
\ee
In addition, we conveniently define the magnetic spectral index $n_B$ via the relation \cite{caprini_2014}
\be
\label{eq:cascade2}
    \dfrac{d\rho^{\pm}_B}{d\ln k} \propto k^{2 n_B}. 
\ee
To test this magnetogenesis scenario against observation, we need to evaluate the magnetic field strength and correlation scale at the end of inflation, being the initial conditions to compute the present values $B_0$ and $L_0$. 
After the end of inflation, the electric field is shorted out due to the high conductivity of the Universe \cite{subramanian_2010}. On the other hand, the evolution of the magnetic field requires accounting for the presence of the cosmic thermal plasma after inflation, which can be described in the Magneto Hydro Dynamic (MHD) limit. It was shown, both analytically and numerically \cite{son_1999, field_2000, christensson_2001}, that helical magnetic fields undergo inverse cascade during the radiation-dominated epoch, leading to an interesting departure from the fully adiabatic evolution in the non-helical case. During inverse cascade, the comoving correlation scale $L_c$ increases, and the comoving field strength $B_c = a^{2} B$ decreases. Power is transferred from small to large scales, while the magnetic spectrum at scales larger than $L$ maintains its spectral index unchanged, displaying a property of self-similarity, namely \cite{caprini_2014}
\be 
\label{eq:Blargescales}
    B(\ell >L) = B \left(\dfrac{L}{\ell}\right)^{n_B}. 
\ee
Following \cite{caprini_2014}, we assume instantaneous reheating after inflation. For a large set of initial conditions, the magnetic field starts evolving in the turbulent regime soon after the end of inflation, conserving comoving magnetic helicity density; after recombination, the magnetic field undergoes adiabatic evolution conserving magnetic flux up to present. Under these conditions, the values of $B_0$ and $L_0$ can be computed from the following relations
\bea
\label{eq:cascade3}
    &B_0 \simeq 10^{-8}\mathrm{G}\left(\dfrac{L_0}{\mathrm{Mpc}}\right),\\
\label{eq:cascade4}
    &B^2_0L_0 = B^2_f L_f\left(\dfrac{a_f}{a_0}\right)^3.
\eea
We should mention that the inverse cascade of magnetic helicity is strongly suppressed if the magnetic field has a (nearly) scale-invariant spectrum, as shown numerically in \cite{brandenburg_2017, kahniashvili_2017}. In this case, to evaluate $B_0$ and $L_0$ it is sufficient to take into account the expansion of the Universe after inflation. 
To evaluate the ratio $(a_f/a_0)$ appearing in equation \eqref{eq:cascade4}, we impose entropy conservation to get the well-known result \cite{subramanian_2016}
\be
\label{eq:ratioscalefactor}
    \dfrac{a_0}{a_f} = \dfrac{g_f^{1/12}}{g_0^{1/3}}\dfrac{\sqrt{H M_{pl}}}{T_0}\left(\dfrac{90}{\pi^2}\right)^{1/4}\sim 0.6\times 10^{29} \sqrt{\dfrac{H}{10^{-5}M_{pl}}}, 
\ee
where $g$ is the effective number of relativistic degrees of freedom and $T$ represents the temperature of the fluid filling the Universe at a given time. We have taken $g_f\sim 100$, $g_0\sim 2.64$ (for two neutrinos species being non-relativistic today), and $T_0 = 2.73$ K.

Note that no assumption has been made so far on the inflationary background evolution. 
The field $\phi$ in the action \eqref{eq:generalaction} can be any scalar field with a non-trivial evolution during inflation, and the associated Lagrangian density $\mathcal{L}_{\phi}$ could be written in principle either in the Jordan or in the Einstein frame. As long as the field $\phi$ can be coupled to the EM sector via $I\left[\phi(a)\right]$ in such a way to reproduce the sawtooth function \eqref{eq:coupling}, the results derived up to here can be applied to any dynamics of inflation. The values of $B_0$ and $L_0$ are entirely determined once the exponents $\nu_i$, the energy scale of inflation, and the total number of e-folds $\Delta N = N_f-N_i$ are specified. In the following, we will apply this mechanism of magnetogenesis to a specific scale-invariant model of modified gravity, where $\phi$ is identified with the inflaton field and the analysis is carried out in the Einstein frame.  
\section{Scale-invariant inflation}
\label{sect:3}
We consider the following quadratic scale-invariant action non minimally coupled to a scalar field, studied in \cite{rinaldi_2016, ghoshal_2022}
\be
\label{eq:RVaction}
    S = \int d^4 x \sqrt{-g}\left[\dfrac{\alpha}{36}R^2 + \dfrac{\xi}{6}\phi^2 R -\dfrac{1}{2}\left(\partial \phi\right)^2 -\dfrac{\lambda}{4}\phi^4\right],
\ee
where $\alpha$, $\xi$, and $\lambda$ are positive and dimensionless arbitrary constants. 

The analysis is simplified if we write the action \eqref{eq:RVaction} in the Einstein frame. Following \cite{rinaldi_2016}, we introduce an auxiliary field $\varphi$ along with an auxiliary variable $\chi$, defined as 
\be
    \chi = \dfrac{\alpha\varphi}{18} +\dfrac{\xi\phi^2}{6}, 
\ee
leading to the following Lagrangian density
\be
\label{eq:L1}
    \dfrac{\mathcal{L}}{\sqrt{-g}} = \chi R -\dfrac{\alpha\varphi^2}{36} - \dfrac{1}{2}\left(\partial \phi\right)^2-\dfrac{\lambda}{4}\phi^4. 
\ee
The equation of motion for the auxiliary field $\varphi$ is $\varphi=R$, making manifest the on-shell equivalence between \eqref{eq:RVaction} and \eqref{eq:L1}. Expressing the Lagrangian in terms of $\chi$ and performing the following Weyl transformation
\be
    \tilde{g}_{\mu \nu}=e^{2\omega(x)}g_{\mu \nu}\ \ \text{where}\ \ \omega =\frac{1}{2}\ln \frac{2\chi}{M^2}\,,
\ee
we find the Einstein frame Lagrangian \footnote{All the quantities in \eqref{eq:LEinst} are evaluated in the Einstein frame, i.e. $R$ and $\sqrt{-g}$ are intended ad the Einstein-frame counterparts of the Jordan-frame quantities appearing in \eqref{eq:L1}.},
\be
\label{eq:LEinst}
    \dfrac{\mathcal{L}_E}{\sqrt{-g}} = \dfrac{M^2}{2}R - \dfrac{3M^2}{\mathfrak{f}^2}\left(\partial \mathfrak{f}\right)^2 - \dfrac{\mathfrak{f}^2}{2M^2}\left(\partial \phi\right)^2 + \dfrac{3\xi\phi^2\mathfrak{f}^2}{2\alpha} - \dfrac{\Omega \phi^4\mathfrak{f}^4}{4\alpha M^4} - \dfrac{9M^4}{4\alpha}, 
\ee
where we have defined $\mathfrak{f} = M e^{-\omega}$ and $\Omega = \alpha\lambda + \xi^2$. $M$ is an arbitrary parameter with mass dimension \footnote{Note that scale invariance is still present at Lagrangian level, as $M$ plays the role of a redundant parameter (see \cite{rinaldi_2016} for a detailed discussion).}. 
Although it might seem that the Lagrangian \eqref{eq:LEinst} gives rise to a two-field inflationary scenario, we can exploit scale invariance to reduce the dynamics to that of a single scalar field \cite{karananas_2016}. The required fields redefinition for this model was found in \cite{tambalo_2017} by explicitly evaluating the Noether's current associated to dilation symmetry, and it reads
\bea
    \label{eq:zeta}& \zeta &= \sqrt{6}M \arcsinh\left(\dfrac{\phi \mathfrak{f}}{\sqrt{6}M^2}\right) ,\\
    \label{eq:rho}& \rho &= \dfrac{M}{2}\ln \left(\dfrac{\phi^2}{2M^2} + \dfrac{3M^2}{\mathfrak{f}^2}\right).
\eea
In terms of the new fields, the Einstein-frame Lagrangian \eqref{eq:LEinst} becomes
\be
\label{eq:LEsinglefield}
    \dfrac{\mathcal{L}_E}{\sqrt{-g}} = \dfrac{M^2}{2}R - \dfrac{1}{2}\left(\partial\zeta\right)^2 - 3\cosh^2\left(\dfrac{\zeta}{\sqrt{6}M}\right)\left(\partial \rho\right)^2 - U(\zeta), 
\ee
where
\be
\label{eq:potential}
    U(\zeta) = -\dfrac{9\xi M^4}{\alpha}\sinh^2\left(\dfrac{\zeta}{\sqrt{6}M}\right) + \dfrac{9\Omega M^4}{\alpha}\sinh^4\left(\dfrac{\zeta}{\sqrt{6}M}\right) + \dfrac{9M^4}{4\alpha}. 
\ee
It is clear that in this representation $\rho$ is the massless mode associated to the flat directions of the potential, i.e. the Goldstone boson. This is consistent with the fact that potential only depends on the field $\zeta$, which is the relevant degree of freedom for the inflationary dynamics. 
The Friedmann equations in a spatially flat FLRW spacetime are 
\bea
    \label{eq:eom1}& H^2 & = \dfrac{\dot{\zeta}^2}{6M^2} + \dfrac{1}{M^2}\cosh^2\left(\dfrac{\zeta}{\sqrt{6}M}\right)\dot{\rho}^2 + \dfrac{U}{3M^2}, \\
    \label{eq:eom2}& \dot{H} & = -\dfrac{\dot{\zeta}^2}{2M^2} - \dfrac{3}{M^2}\cosh^2\left(\dfrac{\zeta}{\sqrt{6}M}\right)\dot{\rho}^2,
\eea
while the Klein-Gordon equations for the two scalar fields are 
\bea
    \label{eq:eom3}&\ddot{\zeta}& + 3H \dot{\zeta} - \dfrac{\sqrt{6}}{2M}\sinh\left(\dfrac{2\zeta}{\sqrt{6}M}\right)\dot{\rho}^2 + \dfrac{dU(\zeta)}{d\zeta} = 0, \\
    \label{eq:eom4}&\ddot{\rho}& + 3H \dot{\rho} + \dfrac{2}{\sqrt{6}M}\tanh\left(\dfrac{\zeta}{\sqrt{6}M}\right)\dot{\rho}\dot{\zeta} = 0.
\eea
As an important remark, once this scale-invariant inflationary background will be coupled to electromagnetism -- as discussed in Sec. \ref{sect:4} -- the Friedmann and Klein-Gordon equations \eqref{eq:eom1}-\eqref{eq:eom4} will carry additional terms. Once proven that EM fields do not back-react, the results for the inflationary dynamics as decoupled from the EM sector discussed here can be safely trusted. 

The analyses performed in \cite{rinaldi_2016, tambalo_2017, vicentini_2019, ghoshal_2022} shows that the system presents an unstable fixed point, corresponding to the local maximum of the potential \eqref{eq:potential}, and a stable fixed point, the minimum. The path from unstable to stable fixed point corresponds to an arbitrarily long quasi-de Sitter expansion phase followed by damped oscillations, with natural identification in inflation and reheating. At the stable configuration, the global scale symmetry gets spontaneously broken and a mass scale is dynamically generated. 
The analysis performed in \cite{ghoshal_2022} shows that a viable inflationary scenario, in agreement with \textit{Planck} constraints on spectral indices \cite{planckcollaboration_2020Inflation}, requires 
\be
\label{eq:bounds}
    \xi^2<\Omega\lesssim \dfrac{2}{\sqrt{3}}\xi^2, \hspace{2cm} \xi \lesssim 1.3 \times10^{-2}. 
\ee
The parameter $\alpha$, which tunes the overall height of the potential $U(\zeta)$, is not constrained from the spectral indices analysis. Nonetheless, knowing that the amplitude of the scalar power spectrum at the pivot scale is \cite{baumann_2009}
\be
    A_S = \dfrac{2}{3\pi^2 M^4_{pl}}\dfrac{U(\zeta_i)}{r}, 
\ee
where $r$ is the tensor-to-scalar ratio, we can set an upper bound on the value of the potential \cite{planckcollaboration_2020Inflation}
\be
    U(\zeta_i) < \left(1.6\times 10^{16}\, \mathrm{GeV}\right)^4, 
\ee
which is equivalent to a lower bound on $\alpha$, 
\be
    \alpha \gtrsim 2\times 10^{10}. 
\ee
For completeness, in Sec. \ref{sect:4} we will verify that the contribution to the tensor-to-scalar ratio from EM fields is negligible. 

\begin{figure}
\centering 
\includegraphics[width=14.5cm]{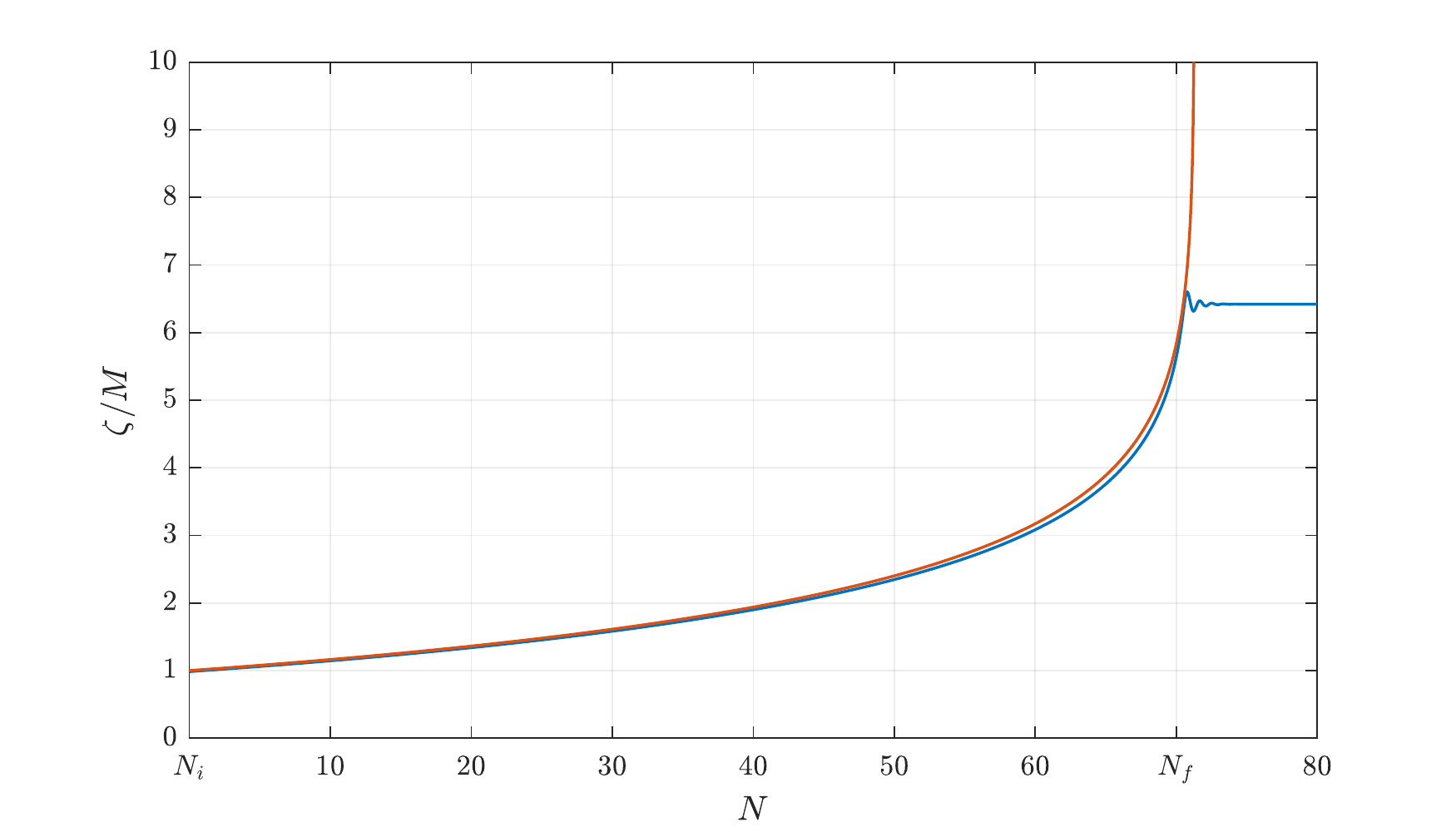}
\caption{Numerical (\textit{blue}) versus approximate (\textit{red}, from equation \eqref{eq:zetaN}) evolution of the inflaton field $\zeta$ in e-folding time $N$. The parameters are fixed as: $\xi = 0.01, \, \Omega= 1.07\, \xi^2, \, \Delta N = 70$.   }
\label{Fig:Zeta_approximation}
\end{figure}

As we will explain in Sec. \ref{sect:4}, to link this inflationary background to the mechanism of magnetogenesis introduced in Sec. \ref{sect:2}, it is useful to find an explicit expression for the evolution of the inflaton field $\zeta$. To this aim, we take the slow-roll limit of equations \eqref{eq:eom1} - \eqref{eq:eom4} and we move to e-folding time $N$, finding the well-known expression \cite{martin_2014}\footnote{We define $N \equiv \ln (a/a_i)>0$}
\be
    N - N_i = -\dfrac{1}{M^2}\int_{\zeta_i}^{\zeta} \dfrac{U(\zeta')}{U_{\zeta'}(\zeta')}d\zeta'. 
\ee
Analytical solutions are found by further expanding the parameters of the model: $\Omega\to\xi^2$ and $\xi\ll1$, in agreement with the bounds \eqref{eq:bounds}. Once the integral has be computed, it can be reversed so to give the approximate expression for the inflaton's evolution,
\be
\label{eq:zetaN}
    \zeta(N) \propto \sqrt{6}M \arctanh\left[\exp\left(\frac{4\xi}{3}N\right)\right], 
\ee
where the proportionality constant is set by initial conditions. In terms of scale factor $a$, 
\be
\label{eq:zetaa}
    \zeta(a) \propto \sqrt{6}M \arctanh\left[a^{\frac{4}{3}\xi}\right]. 
\ee
The validity of the approximation \eqref{eq:zetaN} during the inflationary stage can be appreciated in figure \ref{Fig:Zeta_approximation}. 
\section{Results and discussion}
\label{sect:4}
In the following, we apply the mechanism of magnetogenesis presented in Sec. \ref{sect:2} to the scale-invariant inflationary background discussed in Sec. \ref{sect:3}.
Even though the results of Sec. \ref{sect:2} are rather general, we believe that linking a coupling function $I$ of the form \eqref{eq:coupling} to a specific inflationary dynamics would provide a more robust physical motivation than building $I$ phenomenologically so to obtain the desired magnitude and coherence scale of the generated field. For this reason, we work in the Einstein frame with the fields as defined in \eqref{eq:zeta}-\eqref{eq:rho}, since here the physical content of the model is quite intuitive. Since the Goldstone boson does not take part to the dynamics (equation \eqref{eq:eom4} is trivially solved for $\rho$ = const.), it is natural to couple the inflaton $\zeta$ to the EM field, as first done by Ratra \cite{ratra_1992}.
Having the approximate expression for the inflaton's evolution \eqref{eq:zetaa}, we can derive a coupling function reproducing \eqref{eq:coupling}
\be
\label{eq:couplingzeta(a)}
    I\left[\zeta(a)\right]\sim \tanh\left(\dfrac{\zeta(a)}{\sqrt{6}M}\right)^{\pm \frac{3}{4\xi}\nu_i} \sim a^{\pm \nu_i}, 
\ee
where we omitted the normalization factors. The evolution of the coupling is now related to the evolution of $\zeta$ subjected to the potential \eqref{eq:potential}: $I$ is constant before the field starts its slow-roll motion and becomes again constant when $\zeta$ has settled to the potential's minimum. The full action under consideration is 
\be
\label{eq:completeaction}
    S = -\dfrac{1}{16\pi}\int d^4x\sqrt{-g}\,  I^2[\zeta(a)]\left(F_{\mu\nu}F^{\mu\nu} - \gamma F_{\mu\nu}\tilde{F}^{\mu\nu}\right) + \int d^4x\mathcal{L}_E,
\ee
with $\mathcal{L}_E$ as defined in \eqref{eq:LEsinglefield}-\eqref{eq:potential}. The exponents $\nu_i$ in \eqref{eq:couplingzeta(a)} should be chosen to give a magnetic field in agreement with observational bounds and avoid back-reaction at the same time.  Even though the precise shape of the magnetic power spectrum is not observationally constrained, generating magnetic fields during inflation suggests looking for a (nearly) scale-invariant spectrum, as it happens with scalar and tensor perturbations \cite{mukhanov_2005}. The scale-invariant case is also the most promising one from CMB upper bounds in the perspective of direct detection with next-generation experiments since it is the most weakly constrained \cite{bonvin_2013, subramanian_2016, planckcollaboration_2016, paoletti_2022}. We have seen that for $\nu_1 = 2$ both the magnetic and the electric power spectra are scale-invariant in the first stage of inflation, which avoids any back-reaction. 
The magnetic power spectrum in the second stage, on the other hand, is blue-tilted as $d\rho^{\pm}_{B, II}/d \ln k\sim k^4$. Adding a scale-dependence for $a > a_*$ should not be a problem for CMB bounds since they are only sensitive to the first e-folds of inflation (the CMB observational window encompasses large and intermediate scales, corresponding to a wavenumber range of $10^{-4} \mathrm{Mpc}^{-1}\lesssim k \lesssim 10^{-1} \mathrm{Mpc}^{-1}$ \cite{planckcollaboration_2020Inflation}). The inflationary background we are considering further supports this scenario: we have a scale-invariant model where spontaneous scale-symmetry breaking occurs as the inflaton has settled to the potential's minimum. This is naturally linked to a progressive departure from slow-roll inflation as the steepness of the potential increases. In \cite{tripathy_2022, tripathy_2022_B}, it was shown that such deviations from slow-roll lead to strong features in the power spectra of the EM fields.  
Therefore, it is reasonable to think about a magnetic power spectrum reflecting the scale invariance of the underlying model at the beginning of inflation, which is replaced by a scale-dependent one when the inflaton approaches the end of the plateau and moves towards the minimum. 
We will consider this possibility, unless otherwise stated. 

Before proceeding, we want to make sure that in this scenario there is no back-reaction problem. When the complete action \eqref{eq:completeaction} is considered, the first Friedmann equation \eqref{eq:eom1} takes the form
\be
\label{eq:completeFriedmann}
    3H^2M^2 = \dfrac{1}{2}\dot{\zeta}^2 + 3\cosh^2\left(\dfrac{\zeta}{\sqrt{6}M}\right)\dot{\rho}^2 + U(\zeta) + \rho_{EM}, 
\ee
with $\rho_{EM}$ as defined in \eqref{eq:rhoEM}. The Klein-Gordon equation for the inflaton, equation \eqref{eq:eom3}, becomes
\be
\label{eq:completeKG}
    \ddot{\zeta} + 3H \dot{\zeta} - \dfrac{\sqrt{6}}{2M}\sinh\left(\dfrac{2\zeta}{\sqrt{6}M}\right)\dot{\rho}^2 + \dfrac{dU(\zeta)}{d\zeta} = \mathcal{S}_{EM}
\ee
where we have defined 
\be
\label{eq:SEM}
    \mathcal{S}_{EM} \equiv -\dfrac{1}{8\pi} I \dfrac{dI}{d\zeta}\left(F_{\mu\nu}F^{\mu\nu}-\gamma F_{\mu\nu}\tilde{F}^{\mu\nu}\right). 
\ee
To keep the back-reaction effects under control, we require that the gauge field does not spoil the coupled equations of motion for the inflaton and the background geometry. Consequently, from equations \eqref{eq:completeFriedmann} and \eqref{eq:completeKG}, we must impose two constraints on back-reaction, namely \cite{barnaby_2011, salehian_2021, talebian_2022}
\be
\label{eq:backreaction1}
    \rho_{EM} \ll 3H^2M^2, 
\ee
\be
\label{eq:backreaction2}
    \abs{\mathcal{S}_{EM}} \ll 3 H\dot{\zeta}. 
\ee
Note that we have already accounted for the fact that the terms $\propto\dot{\rho}^2$ are vanishing since $\rho$ is constant.
The two conditions \eqref{eq:backreaction1} and \eqref{eq:backreaction2} must be satisfied throughout inflation, but they are more stringent at the end of inflation, since both $\rho_{EM}$ and $\mathcal{S}_{EM}$ are increasing with time. Note that the constraint \eqref{eq:backreaction2} is stronger than \eqref{eq:backreaction1} by a slow-roll factor $\epsilon$ \cite{talebian_2022}. However, since inflation ends by definition when $\abs{\epsilon} = 1$, it should be sufficient to verify that \eqref{eq:backreaction1} holds to guarantee that back-reaction does not affect the model of magnetogenesis. Nonetheless, we have evaluated both constraints for all the choices of parameters presented below (see figure \ref{Fig:Energies} and figure \ref{Fig:Backreaction2} as an example). 
The expression \eqref{eq:SEM} can be made explicit to simplify the calculation \cite{caprini_2018, salehian_2021}, 
\be
\label{eq:SEM_EB}
    \mathcal{S}_{EM} = \dfrac{1}{4\pi} I \dfrac{dI}{d\zeta} \left(E_iE_i - B_iB_i +2\gamma E_iB_i\right), 
\ee
where $E_i = -a^{-2}A'_i$ and $B_i = a^{-2}\epsilon_{ijk}\partial_jA_k$. In terms of the canonically normalized fields $\mathcal{A_{\pm}}$, we have 
\be
\label{eq:SEM_A}
    \mathcal{S}_{EM} = \dfrac{1}{a^5 \dot{\zeta}} \dfrac{I'}{I}\sum_{(+, -)}\int \dfrac{d^3k}{(2\pi)^3}\left\{ \left|I\left(\dfrac{\A(k, \eta)}{I}\right)'\right|^2-k^2\left|\A\right|^2\mp 2\gamma k \Re\left[\A I\left(\dfrac{\A(k, \eta)}{I}\right)'\right] \right\}, 
\ee
where we have used $dI/d\zeta = \zeta'/a\dot{\zeta}$. The integral \eqref{eq:SEM_A} can be evaluated along the same lines of the magnetic field's amplitude and coherence length, as discussed in Appendix \ref{appendixA}. 

We adopt the following procedure. First, we set the parameters related to the inflationary dynamics: the total number of e-folds $\Delta N$ and the constants $\xi$ and $\Omega$, according to the bounds in equations \eqref{eq:bounds}. In this way, also the values of $\zeta$ at beginning and end of inflation are determined. We then fix $\nu_1 = 2$ to reproduce a scale-invariant magnetic/electric spectrum in the first stage. The remaining parameters are: the exponent $\nu_2$ characterizing the coupling $I$ in the second stage, the energy scale of inflation regulated by the overall height of the inflaton's potential through $\alpha$, and the coupling to helicity $\gamma$ as defined in \eqref{eq:generalaction}. These quantities are varied to avoid back-reaction and provide $(B_0, L_0)$ in agreement with the observational bounds. The results are given in Table \ref{Tab:results}.  
\begin{table}[!ht]
    \centering
    \begin{tabular}{c c c c c c c c}
    \hline
     $\rho^{1/4}_{inf}\, \, [\mathrm{GeV}]\, \, $ & $\quad \Delta N\quad $ & $\quad \nu_1\quad $ & $\quad \nu_2\quad $ & $\quad \gamma\quad $ & $\quad L_0\, \, [\mathrm{Mpc}]\quad  $ & $\quad B_0(L_0)\, \, [\mathrm{nG}]\quad  $ & $B_0(\ell = 1\, \mathrm{Mpc})\, \, [\mathrm{nG}]$\\
      \hline
      \hline
      $3\times 10^{15}$ & 60 & 2 & 1.3 & 1 & 0.74 & 7.4 & 4.0 \\
      $5\times 10^{14}$ & 60 & 2 & 1.4 & 1 & 0.13 & 1.3 & $2.0\times 10^{-2}$ \\
      $3\times 10^{13}$ & 60 & 2 & 1.5 & 1 & $6.7 \times 10^{-3}$ & $6.7\times 10^{-2}$  & $3.0\times 10^{-6}$ \\
      $3\times 10^{13}$ & 70 & 2 & 1.5 & 1 & 0.19 & 1.9 & $6.7\times 10^{-2}$ \\
      $9\times 10^{10}$ & 55 & 1.7 & 1 & 0.4 & $1.3\times 10^{-8}$  & $1.3\times 10^{-7}$ & $1.0\times 10^{-25}$ \\
      $2\times 10^{7}$ & 55 & 3.4 & 0.3 & 0.2 & $4.6\times 10^{-8}$ & $4.6\times 10^{-7}$ & $1.8\times 10^{-11}$\\
      \hline
    \end{tabular}
    \caption{Present day magnetic field amplitude and correlation scale for different choices of the energy scale $\rho^{1/4}_{inf}$ and duration $\Delta N$ of inflation, exponents $\nu_i$, and coupling to helicity $\gamma$. An inverse-cascade evolution has been considered according to equations \eqref{eq:cascade3}-\eqref{eq:cascade4}. The values of $B_0$ and $L_0$ are computed according to the procedure presented in Appendix \ref{appendixA}.}
    \label{Tab:results}
\end{table}

All the listed results for $\nu_1 = 2$ are compatible with the window between gamma-ray lower bounds and CMB upper bounds. At variance with the case of a monotonic coupling $I$ \cite{caprini_2014}, these results are obtained without lowering dramatically $\rho_{inf}$, resulting at the same time in higher $(B_0, L_0)$. We recall here that in many works \cite{martin_2008, ferreira_2013, caprini_2014, sharma_2018} $H$ -- thus $\rho_{inf}$ -- gets lowered to avoid back-reaction since $\rho_{inf}\sim H^2$ and $\rho_{EM}\sim H^4$ in units of the Planck mass. When a sawtooth coupling is considered, the back-reaction problem is alleviated since it only afflicts the second stage of inflation in the cases we have analyzed. This implies an overall higher $\rho_{inf}$, thus a higher maximum $\rho_{EM}$ and, from equation \eqref{eq:cascade1}, a higher $B_f$. It may also happen that the value of $H$ needs to be lowered to keep the tensor-to-scalar ratio $r$ below CMB bounds when the gauge field contributes significantly to the tensor perturbations \cite{caprini_2014}. We will show below that this is not the case for the choice of parameters adopted here.  
At the same time, we have taken advantage of the term $\propto F_{\mu\nu}\tilde{F}^{\mu\nu}$ in the action \eqref{eq:generalaction}, giving rise to fully helical magnetic fields and, consequently, larger values of $L_0$ due to inverse cascade process. This is the main difference from the model discussed in \cite{ferreira_2013}, where no substantial improvement was found by considering a sawtooth coupling in the non-helical case. Note also that, to be consistent with the analysis leading to equations \eqref{eq:cascade3}-\eqref{eq:cascade4}, we have only computed scale-averaged values for the magnetic field amplitude, and not the corresponding fixed-scale values, defined in \eqref{eq:B}.  

\begin{figure}[!ht]
\centering 
\includegraphics[width=14.5cm]{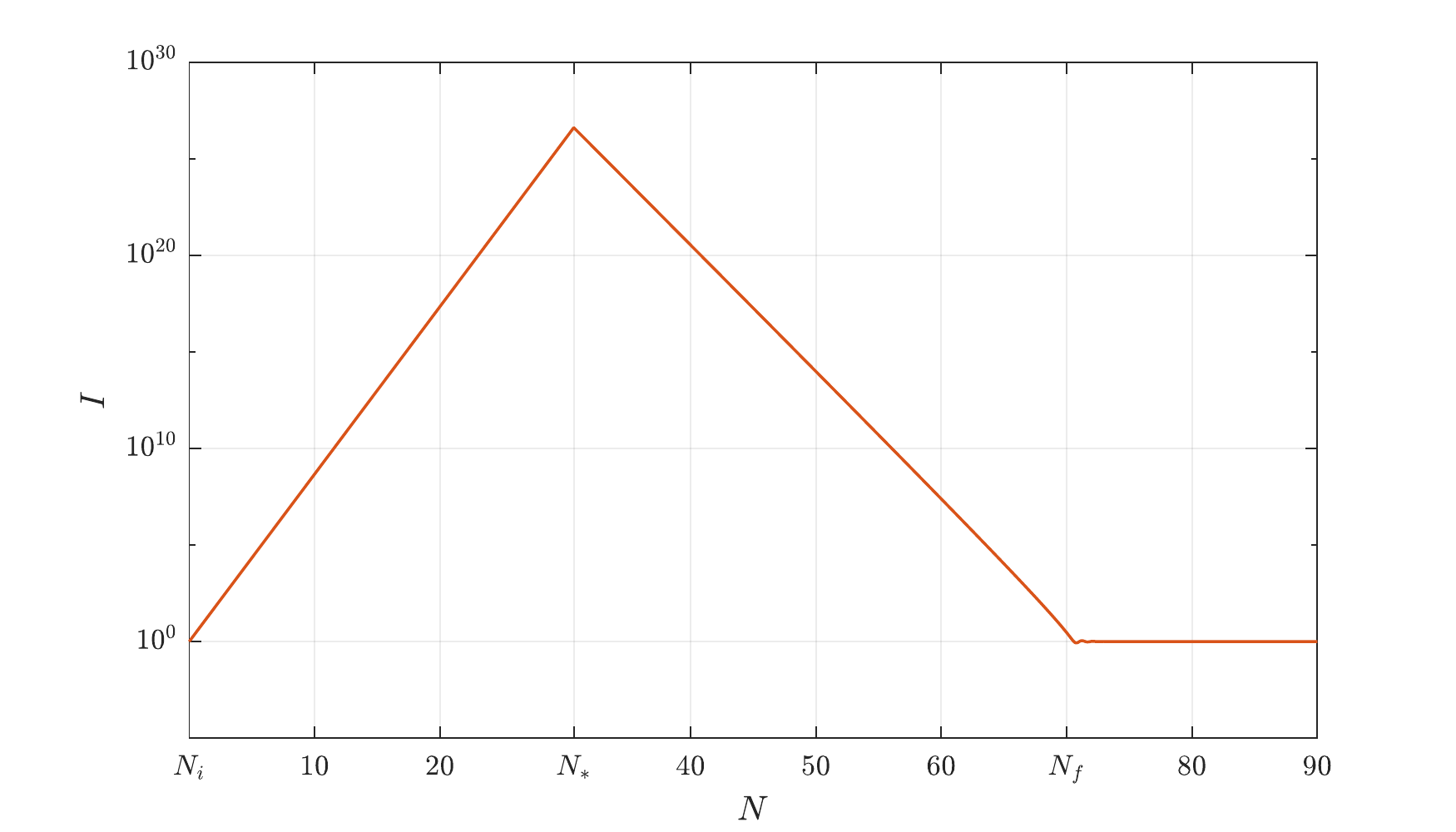}
\caption{Sawtooth coupling during inflation as a function of e-folding time $N$. The parameters are fixed as: $\xi=0.01$, $\Omega=1.07\, \xi^2$, $\Delta N=70$, $\alpha=10^{20}\, \left(\rho^{1/4}_{inf}=3\times 10^{13}\, \mathrm{GeV}\right)$, $\nu_1=2$, $\nu_2=1.5$, $\gamma=1$.}
\label{Fig:Coupling}
\end{figure}

\begin{figure}[!ht]
\centering 
\includegraphics[width=14.5cm]{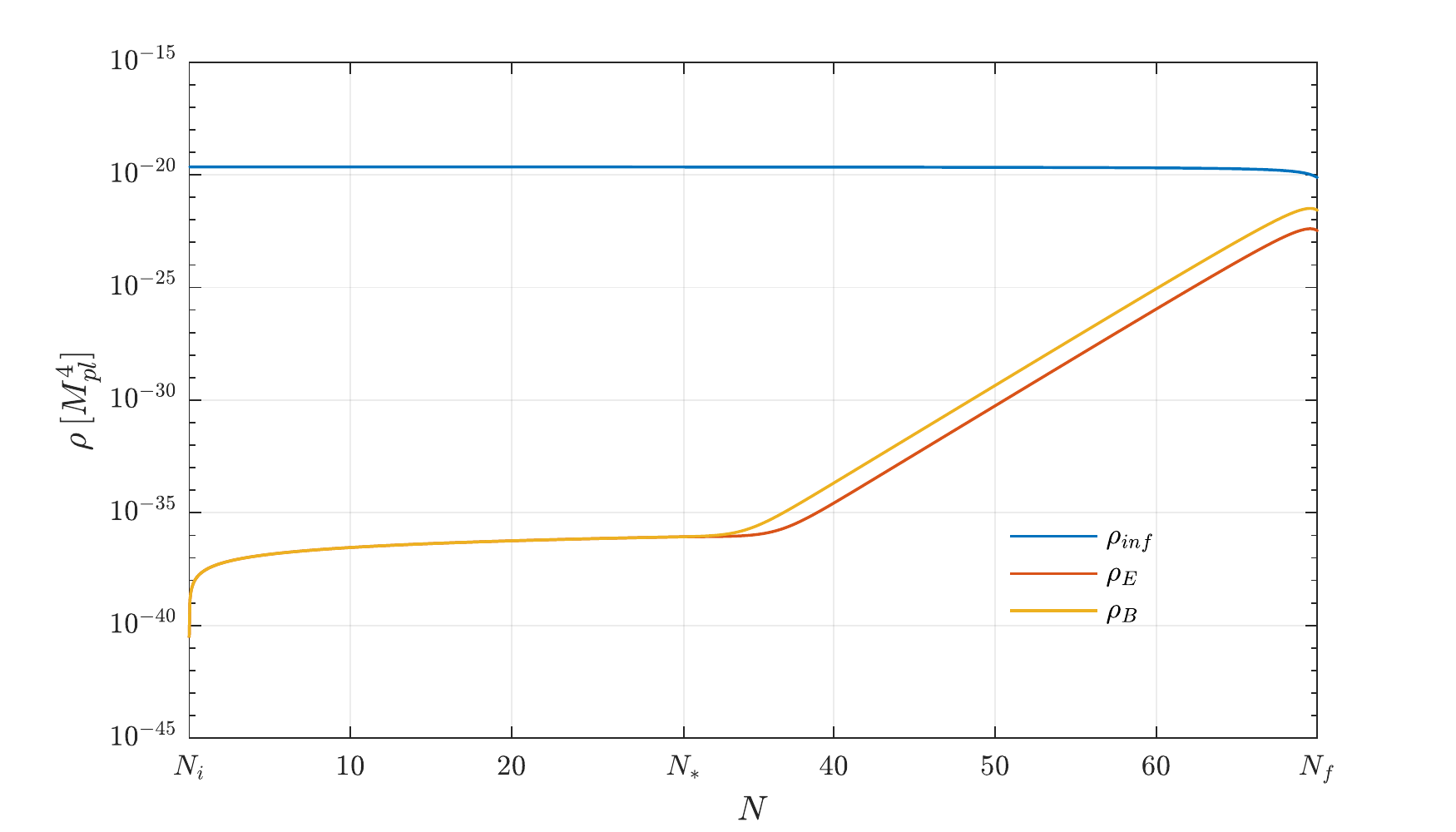}
\caption{Evolution of the inflationary (\textit{blue}) versus electric (\textit{red}) and magnetic (\textit{yellow}) energy density during inflation as a function of e-folding time $N$. The same choice of parameters as in figure \ref{Fig:Coupling} has been made.}
\label{Fig:Energies}
\end{figure}

\begin{figure}[!ht]
\centering 
\includegraphics[width=14.5cm]{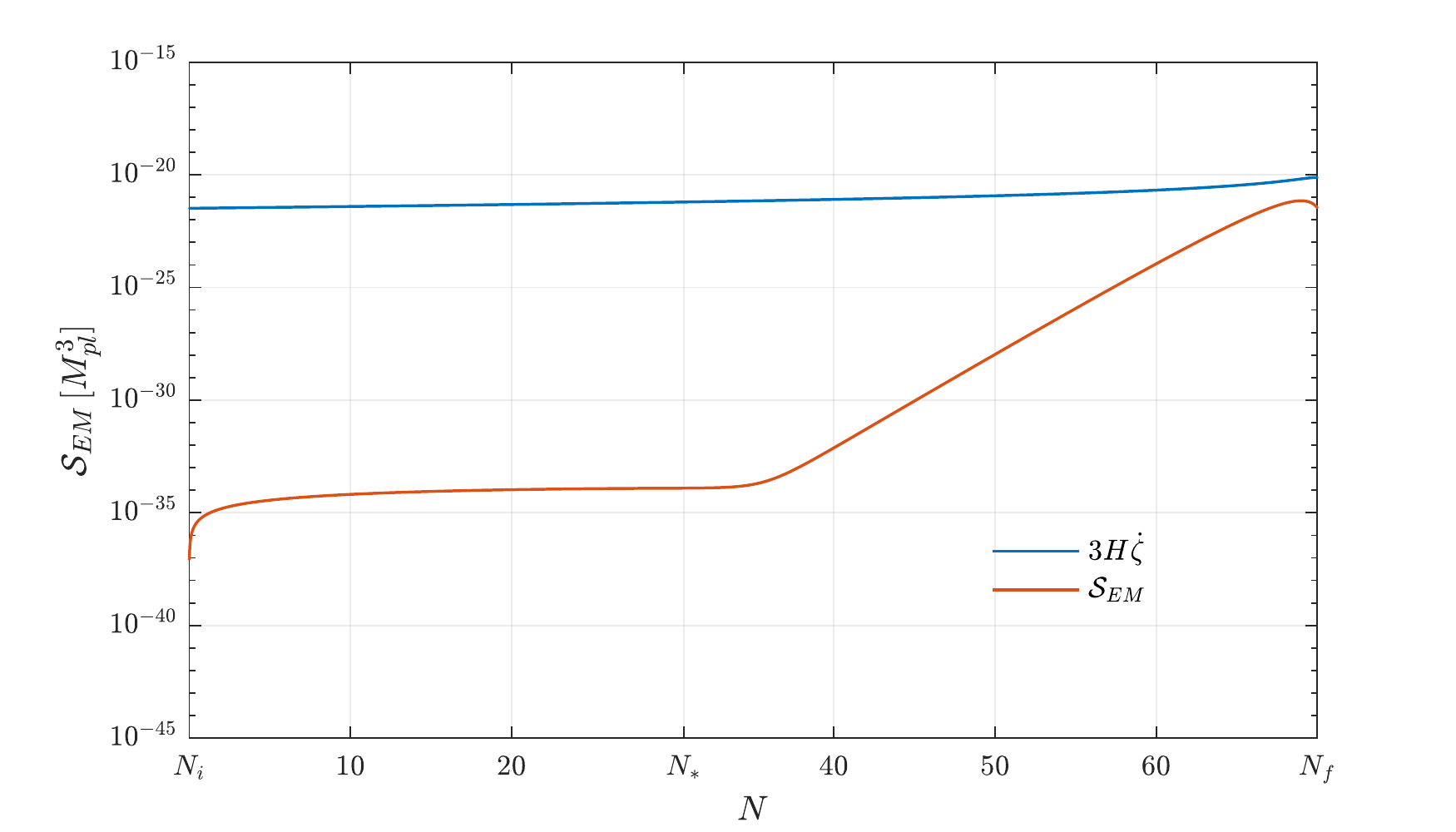}
\caption{Comparison between the right-hand-side (\textit{red}) and the left-hand-side (\textit{blue}) of the constraint equation \eqref{eq:backreaction2}, as a function of e-folding time $N$. The same choice of parameters as in figure \ref{Fig:Coupling} has been made.}
\label{Fig:Backreaction2}
\end{figure}

To test the validity of the slow-roll regime and the approximate result for the coupling function \eqref{eq:coupling}, we support our analysis with numerical solutions to the equations of motion \eqref{eq:eom1}-\eqref{eq:eom4}. Figure \ref{Fig:Coupling} shows the coupling function $I(\zeta)$ using the numerical evolution for $\zeta(N)$. Figure \ref{Fig:Energies} and figure \ref{Fig:Backreaction2} display the two back-reaction constraints \eqref{eq:backreaction1}-\eqref{eq:backreaction2} for one of the above choices for $\nu_i$. 
\subsection{Tensor perturbations sourced by the gauge field}
In the above analysis we have neglected the effect of metric perturbations induced by the gauge field. 
In some models of magnetogenesis \cite{caprini_2014, caprini_2018} it was shown that to have these tensor perturbations under control (i.e., not affecting the properties of the CMB) the energy scale of inflation $\rho_{inf}^{1/4}$ needs to be lowered significantly. In this section, we give a rough estimate of the contribution to the tensor-to-scalar ratio from EM fields. To do so, we employ the same derivation of \cite{caprini_2014, caprini_2018}, originally presented in \cite{sorbo_2011}, even though the calculation performed there refers to a hybrid axion-Ratra model of magnetogenesis with a monotonic decreasing coupling $I$, instead of a sawtooth coupling. Nonetheless, for our purposes it will be sufficient to evaluate the the tensor-to-scalar ratio in the first stage of inflation, where our model matches \cite{caprini_2014} when a monotonic increasing $I$ is considered. As already stated above, \textit{Planck} measurements are only sensitive to the first e-folds of inflation, thus we will not extend our calculation beyond the transition point. By adapting equation (3.7) of \cite{caprini_2014} to the notation adopted here, we have 
\be
\label{eq:boundonr}
    \dfrac{H}{M_{pl}}\dfrac{e^{\pi\xi_1}}{\xi_1^{3/2}} = \left(\dfrac{rA_{\mathrm{S}}}{p^t(-\nu_1)}\right)^{1/4}, 
\ee
where $A_{\mathrm{S}}\simeq 2.1\times 10^{-9}$ is the amplitude of scalar perturbations \cite{planckcollaboration_2020Inflation} and the function $p^t(n)$ is plotted in figure 1 of \cite{caprini_2014}. Given the presence of the exponential function of $\xi_1$ in \eqref{eq:boundonr}, a high-energy inflation is excluded by CMB upper bounds on $r$ when $\xi = \mathcal{O}(10)$, as studied in \cite{caprini_2014}. This is no more true when $\xi < 3$, as already discussed in \cite{talebian_2022}. In the latter case, the usual vacuum tensor perturbations carry the dominant contribution to $r$ and the energy scale of inflation can take higher values. We evaluate $r$ from \eqref{eq:boundonr} by taking the highest energy scale of inflation considered in this paper, 
\be
    \rho_{inf}^{1/4} \simeq 3\times 10^{15}\, \mathrm{GeV} \longrightarrow H \simeq 9\times 10^{-7}\, M_{pl}. 
\ee
We further take $\gamma = 1$ and $\nu_1 = 1.95$\footnote{Note that we have slightly lowered $\nu_1$ from the value adopted above, $\nu_1 = 2$, since the function $p^t(n)$ plotted in \cite{caprini_2014} displays a divergence at $n = 2$. A full derivation for the case of a sawtooth $I$ would be required to conclude that the same divergence plagues also the model discussed here. Even if this is the case, the results in table \ref{Tab:results} would not vary under a change $\nu_1 = 2\to 1.95.$}, so that $\xi_1 \equiv \gamma \nu_1 = 1.95$. We obtain
\be
\label{eq:r}
    r\simeq 7\times 10^{-8}.
\ee
This rough estimate is sufficient to state that the contribution carried from the gauge field to the tensor-to-scalar ratio is negligible if compared to the usual vacuum fluctuations. The latter contribution, as computed for the scale-invariant model \eqref{eq:RVaction} in \cite{rinaldi_2016, tambalo_2017, ghoshal_2022}, is well approximated by the Starobinsky's prediction \cite{starobinsky_1980}
\be
    r \simeq \dfrac{12}{(\Delta N)^2}, 
\ee
which is of the order of $10^{-3}$ for plausible values of $\Delta N = N_f-N_i$. 

\subsection{Baryogenesis from helical magnetic fields}
If helical magnetic fields are produced before the electroweak crossover, they could automatically generate the observed matter/anti-matter asymmetry, without the need of invoking any physics beyond the Standard Model. This issue was discussed in \cite{fujita_2016, kamada_2016, kamada_2016a}.

The conversion from hyper to electromagnetic fields at the EW symmetry breaking causes the decay of hyper magnetic helicity. As a consequence, due to the Standard Model chiral anomaly \cite{thooft_1976}, a (B+L) asymmetry is produced, thus sourcing the baryon asymmetry of the Universe (BAU). It was shown \cite{kamada_2016} that such a generated BAU is not washed out by sphalerons. The observed baryon asymmetry can be explained by hyper-magnetic fields with positive helicity corresponding to intergalactic magnetic fields with $B_0 \sim 10^{-7\sim 8}\,\mathrm{nG}$ and $L_0\sim10^{-2\sim3}\,\mathrm{pc}$.

It would be interesting to see if the mechanism of magnetogenesis described above admits $(B_0, L_0)$ in the right range for baryogenesis since it predicts the generation of magnetic fields with positive helicity. Moreover, the authors in \cite{fujita_2016, kamada_2016, kamada_2016a} adopt the same inverse-cascade evolution we have presented in \ref{sect:2}, thus our results are suitable for a comparison. In the last two lines of Table \ref{Tab:results} we allow for a deviation from scale-invariance in the first stage of inflation by choosing $\nu_1 \neq 2$. By properly varying the parameters and lowering the energy scale of inflation, the model predicts lower values for $(B_0, L_0)$ that are compatible with baryogengesis. As it was already pointed out in \cite{kamada_2016a}, these results are in tension with blazar observations. Higher values of $B_0$ would results in baryon number overproduction. It is possible, however, to choose $\nu_1$ so that $B_0$ is not strongly suppressed on large scales due to formula \eqref{eq:Blargescales}: in the last line of Table \ref{Tab:results}, we show one case where $B_0(\ell = 1\, \mathrm{Mpc})$ is at least large enough to seed dynamo and explain the observed $\mu \mathrm{G}$ magnetic fields in galaxies, requiring minimum magnetic field amplitude in the range \cite{caprini_2014, brandenburg_2005}
\be
10^{-14}\, \mathrm{nG}\lesssim B_0(\ell = 1\, \mathrm{Mpc})\lesssim 10^{-12}\, \mathrm{nG}.
\ee

\section{Conclusions}
\label{sect:5}
In this paper, we have studied the generation of helical magnetic fields during inflation in a modified version of the hybrid axion-Ratra model first presented in \cite{caprini_2014}, where EM conformal invariance is broken only during inflation by a time-dependent function $I(t)$ with a sharp feature. 

We have found that the combination of sawtooth coupling and helical magnetic fields provides current amplitudes and coherence lengths $(B_0, L_0)$ that are sufficient not only to seed galactic fields via cosmic dynamo but also satisfy the lower bounds from blazar observations. 
We have mainly focused on the case of scale-invariant magnetic and electric spectra at large scales/small wavevectors during the first stage of inflation, becoming blue-tilted at smaller scales in the second stage. We have further considered deviations from this picture in the perspective of finding predictions for $(B_0, L_0)$ compatible with the mechanism of baryogenesis first discussed in \cite{kamada_2016}. We found that our model of magnetogenesis is consistent with the observed baryon asymmetry of the Universe and magnetic fields sufficiently strong to seed cosmic dynamo. 

This model is further motivated when a scale-invariant inflationary background is considered. The predictions of magnetogenesis have been computed considering a scale-invariant model of quadratic gravity, as presented in \cite{rinaldi_2016}. In other words, we have cast the evolution of the coupling function to the EM sector into that of the inflaton field, i.e., $I(t) = I[\zeta(t)]$. In this way, we have provided the model with a realistic treatment in a viable inflationary background, which in turn links deviations from scale invariance of the magnetic power spectrum to scale-symmetry breaking characterizing inflation. 
We have checked that the gauge field does not back-react on the inflationary evolution and its contribution to tensor perturbations can be safely neglected.

As a final remark, the results presented in Sec. \ref{sect:2} are applicable to any inflationary model once the total number of e-folds $\Delta N$ and the energy scale $\rho_{inf}$ of inflation are fixed. As a crucial difference from other works \cite{caprini_2014, sharma_2018}, we are not forced to lower the energy scale of inflation significantly.

\appendix
\section{Explicit calculation of magnetic field's properties}\label{appendixA}
In the following, we provide the explicit derivation of the results presented in Table \ref{Tab:results}. 
Since we have considered $I$ to be a piecewise-defined function, also the power spectra are so. Therefore, to compute the energy density stored in the magnetic field as a function of scale factor, following \eqref{eq:rhoEM}, we have 
\be
\label{eq:App_Bdensity}
    \rho^{\pm}_B(a) =
    \begin{cases} 
       \rho^{\pm}_{B, I}(a) = \displaystyle\int_{k_i = a_i H}^{k = aH}\dfrac{dk}{k}\dfrac{d\rho^{\pm}_{B,I}}{d\ln k}& a_i<a<a_*\\
       \displaystyle\rho^{\pm}_{B, I}(a_*) + \int_{k_* = a_*H}^{k = aH}\dfrac{dk}{k}\dfrac{d\rho^{\pm}_{B, II}}{d\ln k}& a_*<a<a_f 
    \end{cases}
\ee
where the magnetic energy densities per logarithmic interval in $k$ are defined in \eqref{eq:PB1} and \eqref{eq:PB2}. All the above integrals can be computed analytically. Note that according to this definition $\rho_B(a)$ is a continuous function of scale factor. 

The physical scale-averaged magnetic field amplitude at the end of inflation is then easily derived according to the definition \eqref{eq:cascade1}, 
\be
\label{eq:App_Bend}
    B_f = \displaystyle\sum_{(+, -)}\sqrt{8\pi \rho^{\pm}_B(a_f)}\, , 
\ee
where we have set $I_f = 1$. Along the same lines, the comoving correlation scale $L_c$ is computed according to \eqref{eq:correlationscale}, 
\be
\label{eq:App_L}
    L_c(a) =
    \begin{cases} 
       L_c^I(a) = \displaystyle\sum_{(+, -)}\dfrac{2\pi}{\rho^{\pm}_{B, I}(a)}\int_{k_i = a_i H}^{k = aH}\dfrac{dk}{k^2}\dfrac{d\rho^{\pm}_{B,I}}{d\ln k}& a_i<a<a_*\\
       \displaystyle L_c^I(a_*) + \sum_{(+, -)}\dfrac{2\pi}{\rho^{\pm}_{B}(a)}\int_{k_* = a_*H}^{k = aH}\dfrac{dk}{k^2}\dfrac{d\rho^{\pm}_{B, II}}{d\ln k}& a_*<a<a_f 
    \end{cases}
\ee
The corresponding physical value computed at the end of inflation is
\be
L_f = a_f L_c(a_f), 
\ee
where we are conventionally adopting $a_f \equiv a_f/a_0$, which can be readily estimated from \eqref{eq:ratioscalefactor}. The resulting value is in units of $M_{pl}^{-1}$. To transform in parsecs we employ the following conversion: $M^{-1}_{pl} = \sqrt{8\pi}\, \ell_{pl} = \sqrt{8\pi}\, (5.2\times10^{-52})$ pc. 
Once the magnetic field's amplitude and correlation scale at the end of inflation are computed, they can be plugged into equations \eqref{eq:cascade3}-\eqref{eq:cascade4} to evaluate the corresponding present-day values after inverse cascade evolution, namely
\be
    B_0 = \left[\left(10^{-8}\,\mathrm{G}\right) B_f^2\left(\dfrac{L_f}{\mathrm{Mpc}}\right)\right]^{1/3}\left(\dfrac{a_f}{a_0}\right), 
\ee
\be
    \dfrac{L_0}{\mathrm{Mpc}} = \dfrac{B_0}{10^{-8}\, \mathrm{G}}. 
\ee
If $L_0 < 1\, \mathrm{Mpc}$, the magnetic field amplitude on Mpc scale is computed following \eqref{eq:Blargescales}, where $n_B$ is taken as the magnetic spectral index in the second stage
\be 
    B_0(\ell = 1\, \mathrm{Mpc}) = B_0\left(\dfrac{L_0}{1\, \mathrm{Mpc}}\right)^{4-\nu_1}, 
\ee
where we have employed the definition \eqref{eq:cascade2} and equation \eqref{eq:PB2}.

\bibliography{ref}{}

\end{document}